\documentclass[fleqn,twoside]{article}
\usepackage{espcrc2}

\usepackage{graphicx}
\usepackage[figuresright]{rotating}
\usepackage{wrapfig}

\def\xssnsn{$\sigma(AuAu \!\rightarrow\! Au^*_{1n} Au^*_{1n} \rho^0) \!=\!{ 2.8 \!\pm\! 0.5\!\pm\! 0.7}$~mb}
\def\xsxnxn{$\sigma(AuAu \!\rightarrow\! Au^*_{xn} Au^*_{xn} \rho^0) \!=\! { 39.7  \!\pm\! 2.8  \!\pm\! 9.7}$~mb}
\def\xsnobrk{$\sigma(AuAu \!\rightarrow\! Au Au \rho^0) \!=\! { 370  \!\pm\! 170 \pm 80}$~mb}

\def\xstot{$\sigma(AuAu \!\rightarrow\! Au^{(*)} Au^{(*)} \rho^0) \!=\! { 460  \!\pm\! 220 \pm 110}$~mb}

\def\ee    {$e^+e^-$}
\def\AuAuee {$AuAu \rightarrow Au^*Au^* + e^+e^-$}

\newcommand{\AmS}{{\protect\the\textfont2
  A\kern-.1667em\lower.5ex\hbox{M}\kern-.125emS}}

\title{ 
Coherent Electromagnetic Processes in Ultra-Peripheral Heavy-Ion Collisions}
\author{F.~Meissner and  V.B.~Morozov for the STAR Collaboration \\
	Lawrence Berkeley National Laboratory, One Cyclotron Rd., Berkeley, CA 94720}
       
\begin{document}

\begin{abstract}
We report measurements for
coherent $\rho^0$ production, $AuAu \!\rightarrow \!AuAu\rho^0$,
and coherent $\rho^0$ and $e^+e^-$ pair  production
accompanied by mutual nuclear Coulomb excitation, $AuAu \!\rightarrow \!Au^\star Au^\star\rho^0$ 
and  \AuAuee, in ultra-peripheral relativistic gold-gold collisions (UPC).
We discuss transverse momentum, mass, and rapidity distributions. 
The two photon-process of $e^+e^-$ pair production is an important probe
of strong field QED because of the large coupling $Z\alpha=0.6$.  At
$\sqrt{s_{NN}}=200$~GeV, the $e^+e^-$ production cross section agrees with lowest
order QED calculations.  The cross sections for coherent $\rho^0$
production at $\sqrt{s_{NN}}=130$ and $200$~GeV are in agreement with
theoretical predictions. The calculations for both, coherent $e^+e^-$ and  $\rho^0$
production treat  nuclear excitation as independent process.
\end{abstract}

\maketitle

In ultra-peripheral heavy ion collisions, photon exchange,
photon-photon or photon-nuclear interactions take place at impact
parameters $b$ larger than twice the nuclear radius $R_A$, where no
nucleon-nucleon collisions occur~\cite{baurrev}.  Examples are nuclear
Coulomb excitation, electron-positron pair and meson production, and
vector meson production.  The exchange bosons can couple coherently to
the nuclei, yielding large cross sections. Coherence restricts the
final states to low transverse momenta, a distinctive experimental
signature.  We report measurements for coherent $e^+e^-$ pair and
$\rho^0$ production, where both processes may be accompanied by mutual
nuclear Coulomb excitation,  $AuAu \!\rightarrow \!Au^{(\star)}
Au^{(\star)}e^+e^-$ and $AuAu \!\rightarrow \!Au^{(\star)}
Au^{(\star)}\rho^0$.

\begin{figure}[!h]
\includegraphics[width=7.5cm,height=2cm,bbllx=0pt,bblly=20pt,bburx=650pt,bbury=180pt]{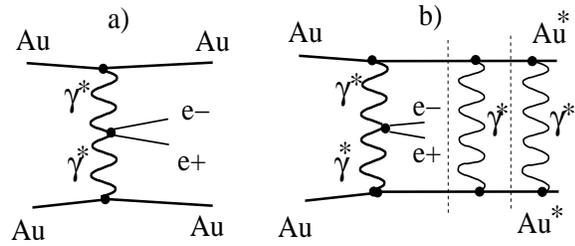}
\caption[]{ Diagram for (a) exclusive $e^+e^-$ pair production in
ultra-peripheral heavy ion collisions, and (b) pair production
with  nuclear excitation. The dashed lines indicate factorization. 
\label{fig:feynman_ee}}
\end{figure}

The purely electromagnetic process of $e^+e^-$ pair production
is shown in lowest order in Fig.~\ref{fig:feynman_ee}a.
The photon flux emitted by the gold ions can be described by the
Weizs\"acker-Williams approach~\cite{weizsaecker}. The $e^+e^-$ pair production 
is then a result of a two-photon collision. 
The gold nuclei are not disrupted, and the final state consists solely
of two nuclei and  two leptons.

The quasi-real photons couple coherently to the ${\rm Au}$ ions; the photon flux
of a single ion is proportional to the square of the nuclear charge
$Z^2$ and the cross-section scales as $Z^4$. The  total cross section
for $e^+e^-$ pair production in ultra-peripheral gold-gold collisions
at $\sqrt{s_{NN}} = 200$~GeV is 33~kb in lowest-order QED.  The applicability of QED
perturbation theory for $e^+e^-$ pair production is questionable since
the photon coupling is large, $Z \alpha \sim 0.6$, and this process
becomes an important test of strong field QED~\cite{baurrev}.

\vspace*{-0.5cm}

In addition to coherent $e^+e^-$ production, the exchange of virtual
photons may excite the nuclei (Fig.~\ref{fig:feynman_ee}b).  In lowest
order, mutual nuclear excitation of heavy ions occurs by the exchange
of two photons~\cite{xsectAuAu,mutualbreakup}. Pair production and
nuclear excitation factorize for heavy-ion collisions, since
non-factorisable diagrams are small~\cite{hencken}.  Because of the
Coulomb barrier for the emission of charged particles, nearly all
nuclear decays following photon absorption include neutron emission~\cite{GDR}. \\

\begin{figure}[!h]
\includegraphics[width=7.5cm,height=2.cm,bbllx=0pt,bblly=20pt,bburx=650pt,bbury=180pt]{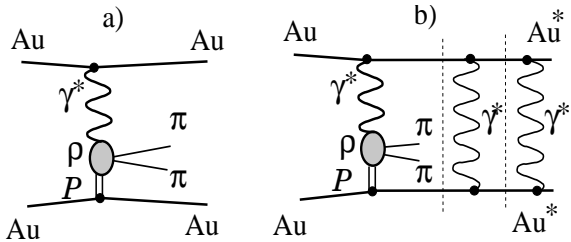}
\caption[]{ Diagram for (a) exclusive $\rho^0$ production in
ultra-peripheral heavy ion collisions, and (b) $\rho^0$ production
with  nuclear excitation. The dashed lines indicate factorization. 
\label{fig:feynman}}
\end{figure}

Exclusive $\rho^0$ meson production, $AuAu\!  \rightarrow\! Au
Au \rho^0$ (Fig.~\ref{fig:feynman}a), can be described by 
the vector meson dominance model~\cite{sakurai}.
A photon emitted by one nucleus fluctuates to a virtual $\rho^0$ meson, 
which scatters elastically from the other
nucleus, yielding a final state of two nuclei and the vector meson decay products~\cite{BKN}.
The additional exchange of virtual photons may  excite the nuclei (Fig.~\ref{fig:feynman}b)
yielding subsequent neutron emission. 
Justified by the similar case of two-photon interactions, both processes are
assumed to factorize.
Calculations for both, coherent $e^+e^-$ and  $\rho^0$
production with nuclear excitation assume that these 
processes are independent, sharing only a common impact
parameter~\cite{BKN,xsectAuAu}.

In the rest frame of the target nucleus, 
mid-rapidity $\rho^0$ production at RHIC  corresponds to  a  photon energy of
$50$~GeV and a photon-nucleon center-of-mass energy of
$10$~GeV. At this energy, Pomeron $(\cal{P})$ exchange dominates over
meson exchange, as indicated by the rise of the $\rho^0$ production cross section with increasing 
energy in lepton-nucleon scattering.
The $\rho^0$ production cross sections are large: 
the photon flux is proportional to $Z^2$~\cite{weizsaecker}, and the forward cross section for
elastic $\rho^0 A$ scattering  $d\sigma^{\rho A}/dt|_{t=0}$  scales as $A^{4/3}$ for surface
coupling and $A^2$ in the bulk limit.  
At a center-of-mass energy of
$\sqrt{s_{NN}}\!=\!130(200)$~GeV, a
total $\rho^0$ cross section, regardless of nuclear excitation,
$\sigma(AuAu
\!\rightarrow\!Au^{(\star)}Au^{(\star)}\rho^0)\!=\!350(590)$~mb is
predicted from a Glauber extrapolation of $\gamma p \!\rightarrow \! 
\rho^0 p$ data~\cite{BKN}. \\

The coherent coupling of photon and Pomeron to the nuclei yields a
unique experimental signature: the central system is created a low
transverse momentum.  For $e^+e^-$ pair production, the
pair-$p_T^{ee}\sim2M_{ee}/\gamma$ is only a few MeV in our accessible
kinematic.
For $\rho^0$ production, the  wavelength $\lambda_{\gamma,\cal{P}}\!>\!2R_A$ leads
to the coherence condition of $p_T \!<\! \pi \hbar/ R_A$ ($\!\sim\!90$~MeV/c for gold with $R_A \!\sim\! 7$~fm)
and a maximum longitudinal momentum of $p_\|\!< \! \pi \hbar \gamma /
R_A$ ($\!\sim\!6(9)$~GeV/c at $\gamma\!=\!70(108)$), where $\gamma$ is the
Lorentz boost of the nucleus.  \\


In the years 2000 and 2001, RHIC collided gold nuclei at a
center-of-mass energy of $\sqrt{s_{NN}}\!=\! 130$ and
$200$~GeV, respectively. The STAR detector consists of a 4.2~m
long cylindrical time projection chamber (TPC) of 2~m radius. In 2000
the TPC was operated in a 0.25~T solenoidal magnetic field. 
In 2001 the magnetic field was at design value of 0.5~T with a 
small data set taken  at 0.25~T. Particles
are identified by their energy loss in the TPC.  A central trigger
barrel (CTB) of $240$ scintillator slats surrounds the TPC. Two zero degree
calorimeters (ZDC) at $\pm$ 18m from the interaction point are
sensitive to the neutral remnants of nuclear break-up.

\section*{$AuAu \rightarrow Au^\star Au^\star e^+e^-$}

Exclusive $e^+e^-$ pair production in UPC has a distinctive
experimental signature: the lepton pair is observed in an otherwise
'empty' spectrometer. The tracks are approximately back-to-back in the
transverse plane due to the small $p_T$ of the pair.  For the analysis
of {$AuAu \rightarrow Au^\star Au^\star e^+e^-$} we use a 800.000
event sample at $\sqrt{s_{NN}}\!=\! 200$~GeV with the  $0.25$~T
magnetic field setting. These events were selected by the 'minimum
bias trigger' which requires coincident neutron signals in the East
and West ZDCs, i.e. mutual nuclear excitation.

\begin{figure}[!h]
\includegraphics[width=7.5cm,height=4.3cm,bbllx=0pt,bblly=60pt,bburx=565pt,bbury=480pt]{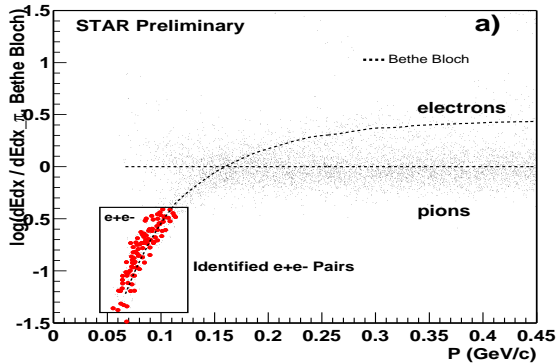} 
\caption{ Energy loss $dE/dx$ of tracks in the 2-track, $\sqrt{s_{NN}}=200$~GeV minimum bias data
with reduced magnetic field. The dots indicate events where both
particles are identified as electrons. \label{fig:electrons} }
\end{figure}

We select events with two oppositely charged tracks, emerging from the
interaction region.  At momenta below $125$ MeV/c, $e^+e^-$ pairs are
identified by their energy loss in the TPC as shown in
Fig.~\ref{fig:electrons}.  To reconstruct a track with at least $90\%$
efficiency requires a minimum track momentum at mid-rapidity ($|\eta
^{track} |\!<\!1.15 $) of $p_{T}^{track}\!>\!65$~MeV/c at $0.25$~T. These requirements
limit the observed \ee pairs to the kinematic range of pair
invariant mass $130 {\rm \ MeV } \!<\! W_{e^+e^-} \!<\! 265 {\rm \ MeV } $,
pair rapidity $ |Y|\!<\!1.15 $ and $\cos(\theta^*) \!<\! 0.7$.  Here,
$\theta^\star$ is the angle between $e^+$ and the photon momentum in
the $\gamma\gamma$ center-of-mass frame (approximately the beam axis).

\begin{figure}[!bth]
\includegraphics[width=4.5cm,height=7cm,angle=270,bbllx=134pt,bblly=30pt,bburx=575pt,bbury=655pt]{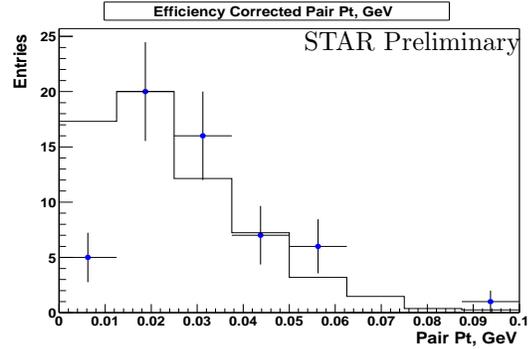}
\caption{\label{pt} 
Efficiency corrected total transverse momentum spectrum of the identified \ee pairs (dots) and scaled Monte Carlo prediction (solid).\vspace*{-0.5cm}}
\vspace*{-6.cm}
\hspace*{4cm}STAR Preliminary
\vspace*{5.5cm}
\end{figure}

We have identified 61 \ee~ultra--peripheral pairs in the event sample
collected with the minimum bias trigger at 0.25T.  The
efficiency corrected $p_{T}^{ee}$  spectrum in Figure
\ref{pt} (dots) shows a clear peak at $p_T \!< \!20$ MeV/c identifying the
process $AuAu \rightarrow Au^\star Au^\star e^+e^-$.  

We have developed a Monte Carlo simulation of the \ee pair production
with mutual nuclear excitation~\cite{star243,morozov}. The photon
fluxes from the ${\rm Au}$ ions are calculated according to the
Weizs\"acker--Williams equivalent photon approximation. The simulation
uses the lowest--order QED approximation for the $\gamma \gamma
\rightarrow e^+e^-$ annihilation and assumes that the pair production
is independent of electromagnetic excitation of the ${\rm Au}$
ions. Except for the lowest $p_T$ bin, the data in Figure~\ref{pt} are in
good agreement with a normalized Monte Carlo simulation (histogram).

\begin{figure}[!bth]
\includegraphics[width=4.5cm,height=7.cm,angle=270,bbllx=134pt,bblly=30pt,bburx=575pt,bbury=655pt]{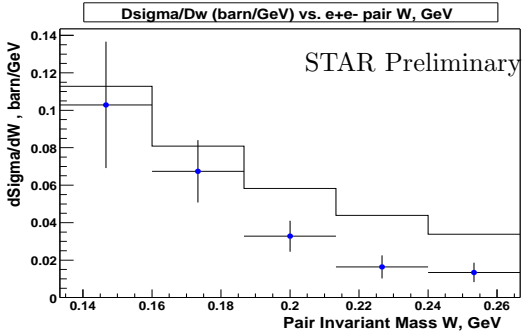}
\caption{\label{minv} Differential cross--section vs. pair invariant mass for the data (dots) and Monte Carlo prediction (solid).\vspace*{-0.5cm}}
\vspace*{-5.5cm}
\hspace*{4cm}STAR Preliminary
\vspace*{5cm}
\end{figure}

The integrated luminosity for the
minimum bias data is determined from the number of hadronic
interactions, assuming a total gold-gold hadronic cross section of
$7.2$~b~\cite{xsectAuAu}.
The differential cross--section $d\sigma^{ee}/dW$ is presented in Figure \ref{minv}, 
compared to Monte Carlo simulations. Within the limited kinematic range quoted above
the integrated  cross--section for coherent electron pair production was found to be
$6.0$~mb$\pm 17\% \pm 19\%$ which agrees well with the Monte Carlo
prediction of $7.8$~mb.
The major contributions for the systematic uncertainty come from the luminosity 
determination, efficiency corrections for  tracking and  vertex finding,  and acceptance correction.

The cross section $d\sigma(AuAu\rightarrow
e^+e^-)/dW_{\gamma\gamma}$ is maximal at low invariant mass
$W_{\gamma\gamma}\!\sim\!2m_e$ and falls rapidly as
$1/W_{\gamma\gamma}^4$; the pair production is peaked at small forward
angles $\theta^\star\!\sim\!1/\gamma$. Including the requirement of nuclear breakup, only a small
($10^{-6}$) fraction of the total cross section for coherent lepton
pair production is observable. Even at the
$0.25$~T field, low momentum electrons do not reach the central
trigger barrel, and triggering is limited to the minimum bias trigger,
i.e. $e^+e^-$ pair production with mutual nuclear excitation.
A larger kinematic acceptance for \ee~pair production could be accomplished by
lowering the field of the STAR magnet.
In the limited kinematic range and limited by the available statistics, 
our measurement agrees with leading order QED calculations. 
Extending these measurements to lower values of $W_{\gamma\gamma}$ would 
allow to access the kinematic region, where the contributions of higher order effects 
are expected to become  sizable.

\section*{$Au Au\!\rightarrow\! AuAu\rho^0$ and $Au Au
\!\rightarrow\! Au^\star Au^\star \rho^0$}

Two different triggers are used for the $\rho^0$ analysis.  For $AuAu\! 
\rightarrow\! AuAu \rho^0$, about 30,000 (2000) and 1.5~M(2001) events
were collected using a low-multiplicity `topology' trigger.  The CTB
was divided in four azimuthal quadrants. Single hits were required in
the opposite side quadrants; the top and bottom quadrants acted as
vetoes to suppress cosmic rays.  
A fast on-line reconstruction removed events without reconstructible tracks
from the data stream.  To study $Au Au
\!\rightarrow\! Au^\star Au^\star \rho^0$, about 0.8~M(2000) and 2.5~M(2001) 
`minimum bias' events, which required coincident detection of neutrons
from nuclear break-up in both ZDCs as a trigger, are used for the
analysis.

\begin{figure}
\includegraphics[width=5.7cm,height=4.2cm]{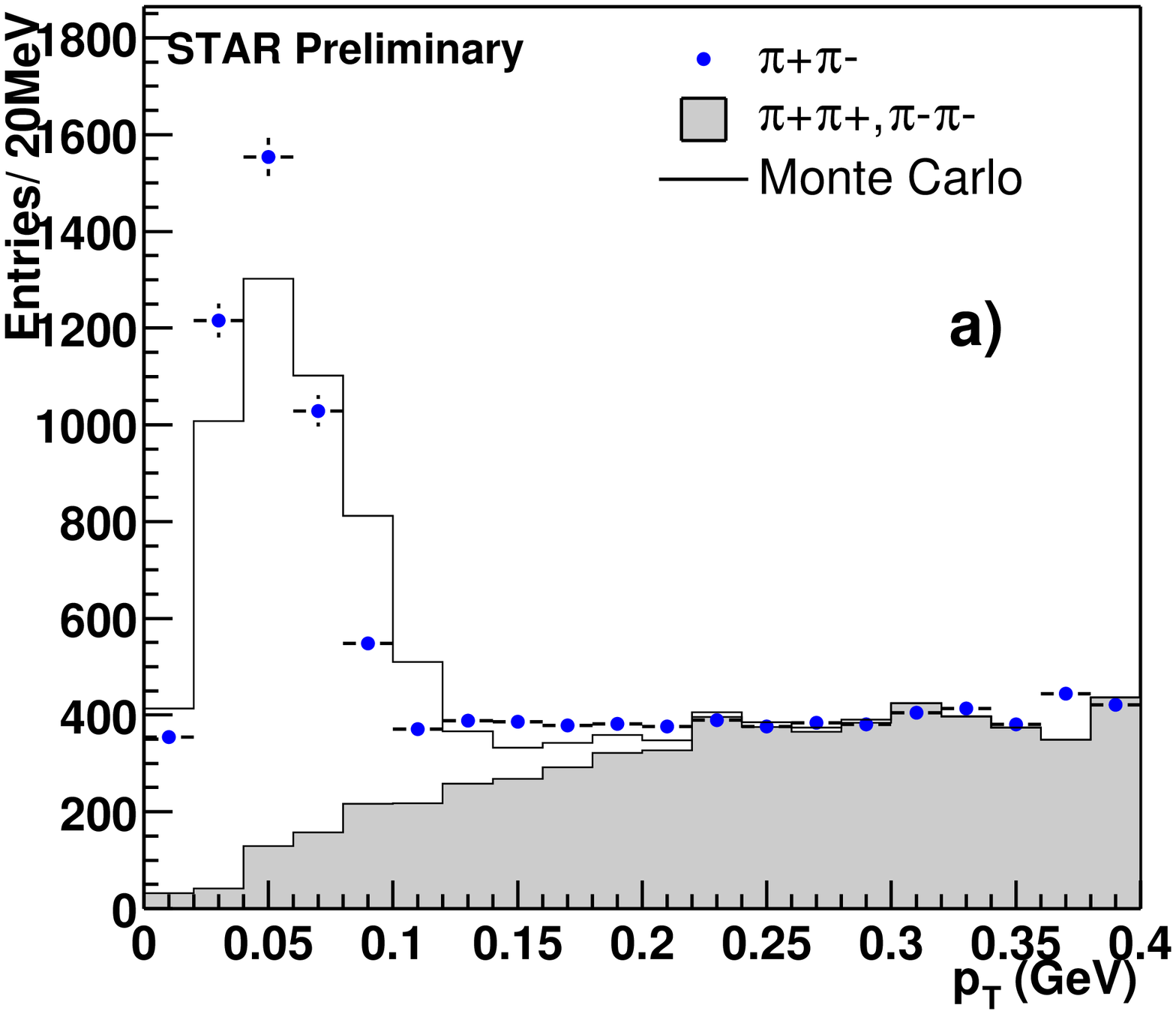} 
\includegraphics[width=5.7cm,height=4.2cm]{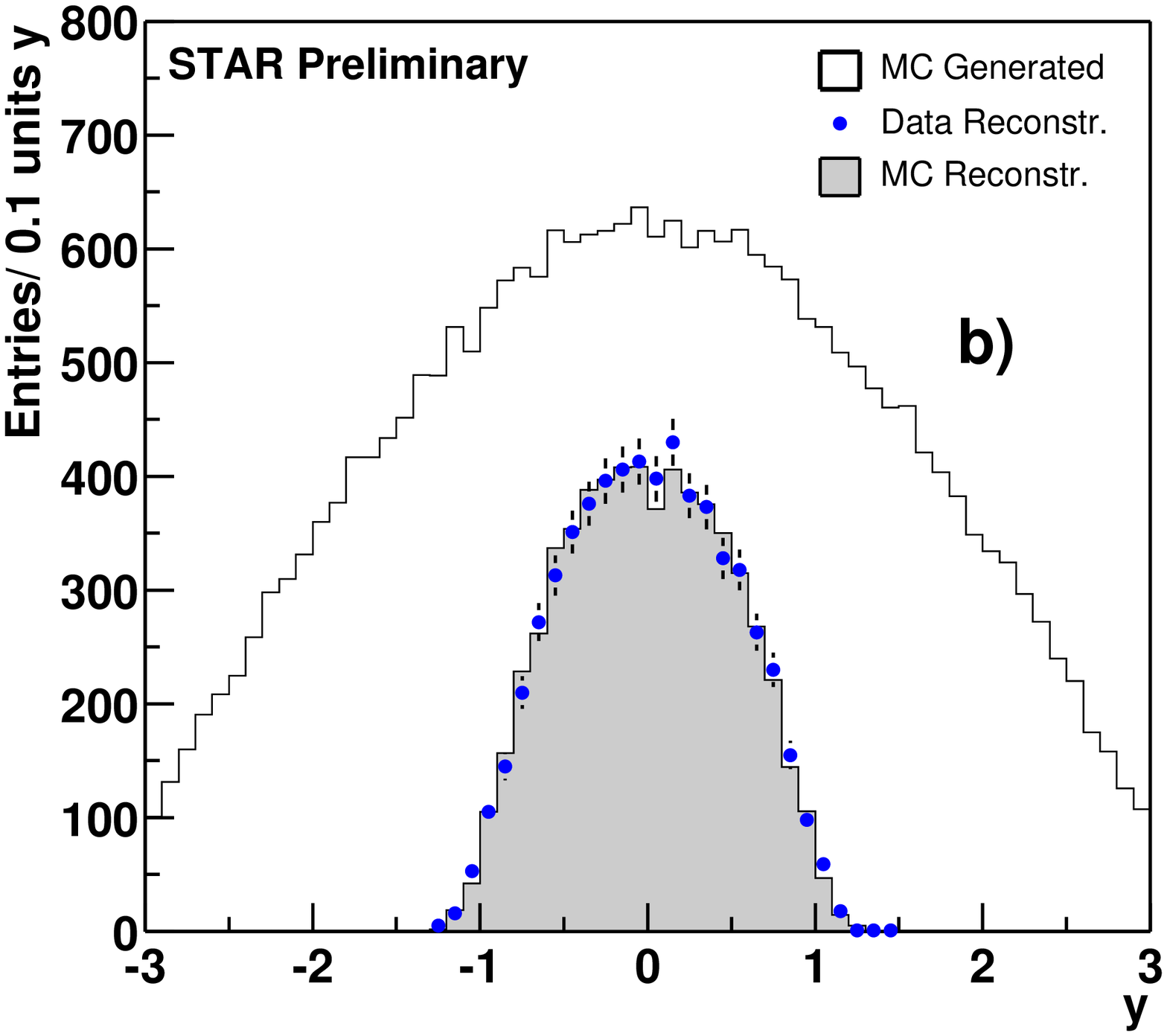}
\includegraphics[width=5.9cm,height=4.2cm]{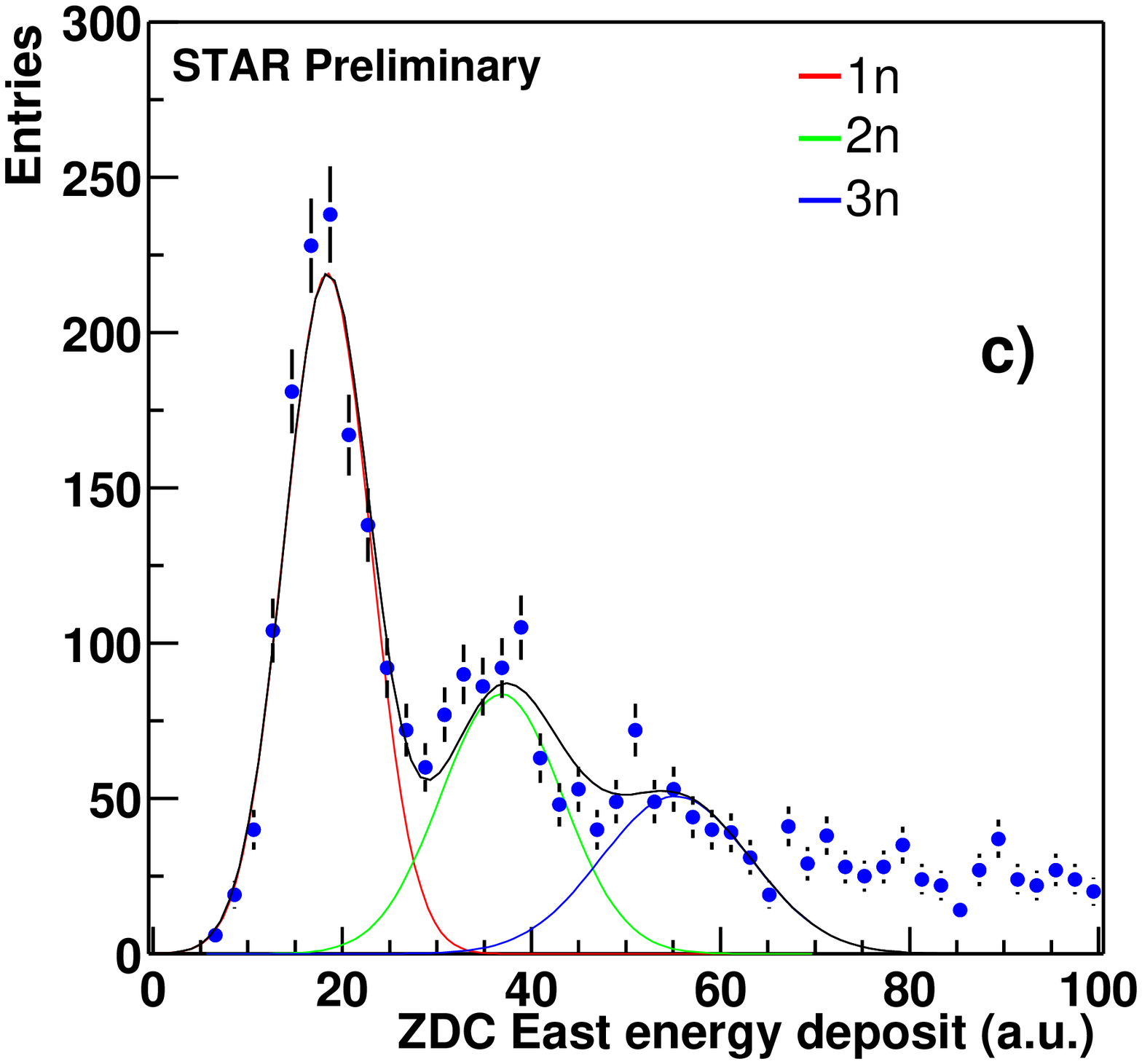}
\includegraphics[width=5.7cm,height=4.2cm]{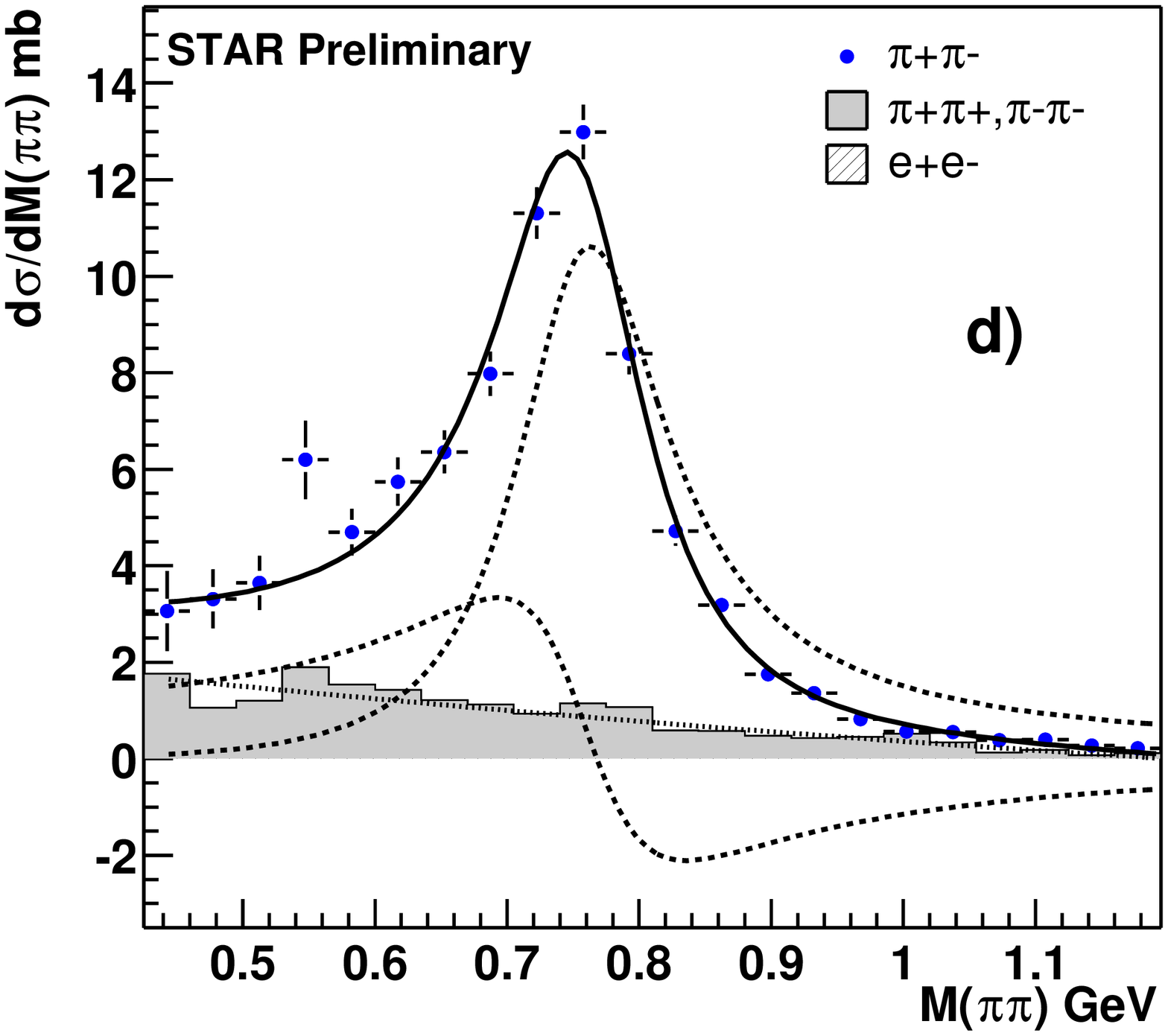}
\caption{
The (a) $\rho^0$ transverse momentum  and (b) rapidity distribution,
the (c) ZDC response, and (d) the $d\sigma/dM_{\pi\pi}$ invariant mass
distribution for  2-track (xn,xn) events in the $\sqrt{s_{NN}}\!=\!200$~GeV minimum bias data. 
\label{fig:all}}
\end{figure}

Events are selected with exactly two oppositely charged tracks forming
a common vertex within the interaction region.  The specific energy
loss $dE/dx$ in the TPC shows that the event sample is dominated by
pion pairs. In the topology triggered data sets, without the ZDC
requirement, cosmic rays are a major background.  They are removed by
requiring that the two pion tracks have an opening angle of less than
3 radians.
Figure~\ref{fig:all} shows kinematic distributions for the selected
2-track events in the $\sqrt{s_{NN}}\!=\!200$~GeV minimum bias data; these
distributions  are similar for the other data sets.

Figure~\ref{fig:all}a) shows the transverse momentum spectrum of
oppositely charged pion--pairs (points).  A clear peak, the signature
for coherent coupling, can be observed at $p_T\!<\!150$~MeV/c. Those
events are compatible with coherently produced $\rho^0$ candidates.  A
background model from like-sign combination pairs (shaded histogram),
which is normalized to the signal at $ p_T \!>\!$ 250 MeV/c, does not
show such a peak.  The open histogram is a Monte Carlo
simulation~\cite{BKN} for coherent $\rho^0$ production accompanied by
nuclear break-up superimposed onto the background.  The $dN^\rho/dp_T$
(i.e. the $dN^\rho/dt \!\sim \!dN^\rho/dp_T^2$) spectrum reflects not
only the nuclear form factor, but also the photon $p_T$ distribution
and the interference of production amplitudes from both gold
nuclei. The interference arises since both nuclei can be either the
photon source or the scattering target~\cite{vminterf}.  A
detailed analysis of the $p_T$ distribution is in progress.

The rapidity distribution in Fig.~\ref{fig:all}b) is well
described by the reconstructed events from the Monte Carlo simulation.
The generated rapidity distribution is also shown.  The acceptance
for exclusive $\rho^0$ is about $40\%$ at $|y_\rho|\!<\!1$.
At $|y_\rho|\!>\!1$, the acceptance is small and this region is excluded from the analysis; the cross
sections are extrapolated to the full $4\pi$ acceptance with the Monte Carlo simulation.
Using the energy deposits in the ZDCs (Fig.~\ref{fig:all}c), we select
events with at least one neutron (xn,xn), exactly one neutron (1n,1n),
or no neutrons (0n,0n) in each ZDC; the latter occurs only in the
topology trigger.

Figure~\ref{fig:all}d) shows the $d\sigma/dM_{\pi\pi}$ spectrum for
events with pair-$p_T\!< \!150$~MeV/c (points).  The fit (solid) is
the sum of a relativistic Breit-Wigner for $\rho^0$ production and a
S\"oding interference term for direct $\pi^+\pi^-$
production~\cite{soeding} (both dashed). A second order polynomial
(dash-dotted) describes the combinatorial background (shaded
histogram) from grazing nuclear collisions and incoherent
photon-nucleon interactions. The $\rho^0$ mass and width are
consistent with accepted values~\cite{PDG}. Alternative parameterizations like
a modified S\"oding parametrization~\cite{ZEUS} and a
phenomenological Ross-Stodolsky parametrization ~\cite{rosssto} yield similar results.
Incoherent $\rho^0$ production, where a
photon interacts with a single nucleon, yields high $p_T$ $\rho^0$s,
which are suppressed by the low pair $p_T$ requirement; the remaining
small contribution is indistinguishable from the coherent process.  A
coherently produced background arises from the two-photon process
$AuAu\! \rightarrow\! Au^{(\star)} Au^{(\star)} l^+l^-$. It
contributes mainly at low invariant mass $M_{\pi\pi} \!<\! 
0.5$~GeV/c$^2$. The small contribution from $\omega$ decays is
neglected.

\begin{figure}[!h]
\includegraphics[width=7.5cm,height=4cm,bbllx=0pt,bblly=30pt,bburx=520pt,bbury=360pt]{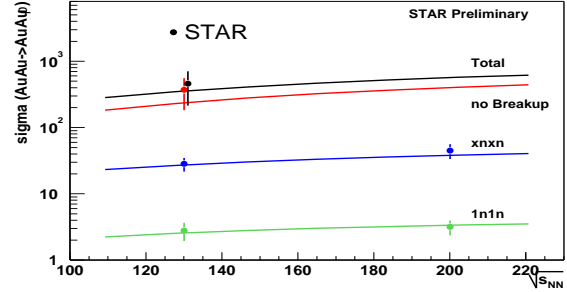}
\caption{Comparison to predictions from Ref.~\cite{BKN}. 
\label{fig:comp} \vspace*{-0.5cm}}
\end{figure}

Figure~\ref{fig:comp} compares our results on cross sections for
coherent $\rho^0$ production at $\sqrt{s_{NN}}=130$~GeV\cite{starrho}
to the calculations of Ref.~\cite{BKN}. Preliminary results for
$\sqrt{s_{NN}}=200$~GeV are also shown in the plot.  The cross
sections are obtained from the integral of the Breit-Wigner fit,
extrapolated to full rapidity.  For coherent $\rho^0$ production
accompanied by mutual nuclear break-up (xn,xn), we measure a cross
section of \xsxnxn.  By selecting single neutron signals in both ZDCs,
we obtain \xssnsn.  The systematic uncertainties are dominated by the
uncertainties of the luminosity determination and the $4\pi$
extrapolation.  The absolute efficiency of the topology trigger is
poorly known and does not allow a direct cross section measurement.
From $\sigma(AuAu\!\rightarrow\! Au^*_{xn} Au^*_{xn}
\rho^0)$ and the ratio $\sigma^\rho_{xn,xn}/\sigma^\rho_{0n,0n}$ we
estimate \xsnobrk~  and the total cross section for coherent $\rho^0$ production \xstot.
From a fit to the differential cross section $d\sigma(\gamma Au\!\rightarrow\!\rho
Au)/dt \!\sim \!d\sigma(\gamma Au\!\rightarrow\!\rho Au)/dp_T^2 \propto e^{-bt}$ for
the (xn,xn) events  we 
obtain a forward cross section $d\sigma^{\rho
A}/dt|_{t=0}\!=\!965\pm140\!\pm\!230$ mb/GeV$^2$ and an approximate
gold radius of $R_{Au}\!=\!\sqrt{4b}\!=\!7.5\!\pm\!2$~fm, comparable
to previous results~\cite{alvensleben}.


\begin{figure}[!t]
\includegraphics[width=7.cm,height=4.2cm]{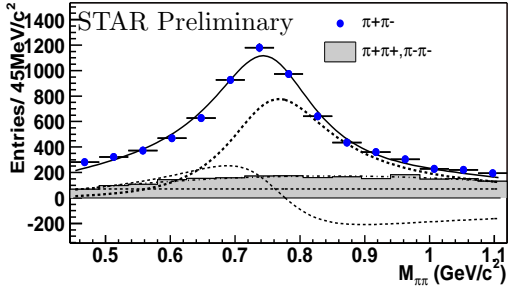}
\caption[]{ The $dN/dM_{\pi\pi}$ invariant mass
distribution for  2-track events in $\sqrt{s_{NN}}\!=\!200$~GeV deuterium-gold collisions.
\label{fig:rapidity}}
\vspace*{-6.1cm}
\hspace*{1cm}STAR Preliminary
\vspace*{5.5cm}
\end{figure}

In the year 2003, STAR recorded deuterium-gold (dAu) collisions. In $dAu$ collisions two processes compete, 
where either  gold or the deuterium are the photon emitter: $Au\rightarrow\gamma Au; \gamma d\rightarrow d\rho$ and 
 $d\rightarrow\gamma d; \gamma Au\rightarrow Au\rho$.  About 1M $dAu$ collisions where recorded 
using the low multiplicity topology trigger, with a sub-sample that included a requirement on a neutron signal in the ZDC, 
i.e. deuterium break up.  Fig.~\ref{fig:rapidity} shows a clear $\rho^0$ signal in 
a preliminary $M{\pi\pi}$ invariant mass spectrum.

In summary, ultra-peripheral heavy-ion collisions are a new
laboratory to study purely electromagnetic and 
diffractive interactions.
The cross section for the two-photon process 
\AuAuee agrees with leading order QED calculations in the limited kinematic 
range presently experimentally accessible.
The first measurements of coherent $\rho^0$ production
with and without accompanying nuclear excitation, $AuAu
\!\rightarrow\! Au^\star Au^\star \rho^0$ and $AuAu \!\rightarrow\! Au
Au \rho^0$, confirm the existence of vector meson production in
ultra-peripheral heavy ion collisions.
The cross sections at $\sqrt{s_{NN}}\!=\!130$ and $200$~GeV are in
agreement with theoretical calculations.
The calculations for both, coherent $e^+e^-$ and  $\rho^0$
production treat nuclear excitation as independent process.

\end{document}